\documentstyle[sprocl]{article}

\newcommand{\ket}[1]{\left | \, #1 \right \rangle}
\newcommand{\bra}[1]{\left \langle #1 \, \right |}

\def\Id{{\mathbf 1}}
\title{Holonomic quantum logic gates}
\author{Marie Ericsson}
\address{Department of Quantum Chemistry, \\
Uppsala University, Box 518, Se-751 20 Uppsala, Sweden\\
Email:marie.ericsson@kvac.uu.se}
\begin{document}
\maketitle \abstracts{This is a brief overview of quantum
holonomies in the context of quantum computation. We choose an
adequate set of quantum logic gates, namely, a phase gate, the
Hadamard gate, and a conditional-phase gate and show how they can
be implemented by purely geometric means. Such gates may be more
resilient to certain types of errors.}

\section{Introduction}
Any quantum computation can be build out of simple operations
involving only one or two quantum bits (qubits). Such operations
are called quantum logic gates and a finite number of them
suffices to construct any quantum boolean network, and therefore
any quantum computer. Moreover, efficient and reliable
quantum-coherent computation of arbitrarily long duration is
possible, even with faulty components. That is, errors can be
corrected faster than they occur, even if the error correction
machinery is faulty. Such fault tolerance is only possible if the
size of the systematic error in the quantum logic gates is not
greater than $10^{-8}$ (possibly a pessimistic estimate).
Motivated by this precision requirement we here consider quantum 
logic gates which are of a geometric / holonomic rather than
dynamical origin and therefore are resilient to certain types of
errors. They may offer the potential of an intrinsically
fault-tolerant computation.

As a set of adequate (universal) quantum gates we choose a phase
gate, the Hadamard gate, and a conditional-phase gate. In the
standard notation, in which the computational basis is formed out
of the two orthogonal states of a quantum bit (qubit) labelled as
$\ket{0}$ and $\ket{1}$, the action of the gates is defined as
follows:

\begin{itemize}
\item The phase gate: $\ket{0} \mapsto
\ket{0}$ and $\ket{1}\mapsto e^{i\alpha }\ket{1}$, for some
prescribed $\alpha$. \item  The Hadamard gate: $ \ket{0} \mapsto
\frac{1}{\sqrt 2} (\ket{0}+ \ket{1})$, $\ket{1} \mapsto
\frac{1}{\sqrt 2} (\ket{0}- \ket{1})$.
\item Two-qubit conditional phase gate: $\ket{00}\mapsto\ket{00}$,
$\ket{01}\mapsto\ket{01}$, $\ket{10}\mapsto\ket{10}$,
$\ket{11}\mapsto e^{i\beta}\ket{11}$, again for some prescribed
$\beta$.
\end{itemize}

In the following we provide a very brief description of the theory
behind holonomic quantum logic gates.

\section{Abelian and Non-Abelian Holonomy}
If an energy eigenstate of a quantal system depends on a
set of external parameters $\lambda=\{\lambda_1,\cdot\cdot\cdot, \lambda_n\}$ 
then an adiabatic cyclic variation of these parameters returns the 
system to its original state. The final state vector turns out to be 
related to the initial state vector via the product of a 
dynamical phase factor, which depends on the speed of the curve
in the parameter space, and a 
geometric phase factor~\cite{berry}
$\exp{\left(i\gamma\right)}$, which depends only on the shape of
path in parameter space,
\begin{equation}
\ket{\psi (t)}= \exp{\left(i\gamma\right)}\;\ket{\psi (0)}
\exp{\left[-\frac{i}{\hbar}\int_0^t E(\tau)\,d\tau\right]}.
\end{equation}

If the state $\ket{\psi }$ belongs to a degenerate subspace it
remains in the degenerate subspace during an adiabatic evolution. 
However, the system returns in general to 
a final state $\ket{\psi (t)}$ related to the initial state 
$\ket{\psi (0)}$ by some unitary operator $U$
\begin{equation}
\ket{\psi (t)}=U \ket{\psi (0)}
\exp{\left[-\frac{i}{\hbar}\int_0^t E(\tau)\,d\tau\right]},
\end{equation}
where $U$ depends only on the shape of the path in parameter
space. Matrix $U$ is a generalization of the geometric phase
(multiplication by the $\exp{\left(i\gamma\right)}$ factor) into
non-Abelian cases~\cite{zee}.

In order to evaluate $U$ let us choose $N$-fold energy degenerate
reference states $\ket{n(\lambda)}$, being local in parameter space. 
Then at any point on the adiabatic path in the parameter
space $\lambda(\tau)$ we can write
\begin{equation}
\ket{\psi_l(\tau)}= \sum_n U_{ln}(\tau) \ket{n(\lambda(\tau))}.
\end{equation}
In particular, for a closed loop the final state
$\ket{\psi_l(t)}=\sum_n U_{ln}(t) \ket{n}$ is related to the
initial state, chosen to be $\ket{l}$, via $U_{ln}(t)$. These
matrix elements can be evaluated as the path ordered line
integral
\begin{equation}
U=\mbox{\bf P} \exp\left[ i\int_0^t
A_{\lambda}\frac{d\lambda}{d\tau} d\tau\right]
\end{equation}
with the gauge potential defined as
\begin{equation}
A_{ab,\lambda}=i\bra{a}\frac{\partial}{\partial\lambda}\ket{b}.
\end{equation}
In the Abelian case, i.e. the geometric phase, the path order is
not necessary and the integral reduces to a regular linear
integral.

\section{Holonomic phase gate}
The simplest case is a single qubit phase gate. The
reference state $\ket{n(\theta,\phi)}$ can be written as
\begin{equation}
|n(\theta,\phi)\rangle=\cos\frac{\theta}{2}|0 \rangle
+e^{i\phi}\sin\frac{\theta}{2}|1 \rangle, \label{eq:state}
\end{equation}
where $\lambda=\{\theta,\phi\}$ are the spherical polar angles of
the Bloch vector. The gauge potential reads
\begin{eqnarray}
A_{\theta}&=&i\bra{n}\frac{\partial}{\partial\theta}\ket{n}=0,\\
A_{\phi}&=&i\bra{n}\frac{\partial}{\partial\phi}\ket{n}=-\frac{1}{2}
\left(1-\cos\theta\right).
\end{eqnarray}

Suppose that a qubit, for example a spin half nucleus in a slowly
varying magnetic field, undergoes a cyclic conical evolution with cone
angle $\theta$. Then the line integral of the gauge potential
gives geometric phase $\gamma
=\pm\frac{1}{2}\Omega=\pm\pi(1-\cos\theta)$, where the $\pm$ signs
depend on whether the system is in the eigenstate aligned with or
against the field, and $\Omega$ is the solid angle subtended by
the conical circuit. Thus the two qubit states $\ket{0}$ and
$\ket{1}$ may end up with geometric phases of the opposite sign,
which gives a phase gate with shift $2\gamma$ between the two
states.

The most common experimental realization is a qubit coupled to an
oscillating electromagnetic field. If $\omega_{0}$ is the
transition frequency of the qubit, $\omega$ is the frequency of
the oscillating field, and $\omega_{1}$ is the amplitude of the
oscillating field then by controlling $\omega$ and $\omega_1$ one
can effectively implement the conical circuit equivalent to that
of slowly varying magnetic field with $\theta$ given by
\begin{equation}\label{costheta}
\cos\theta=\frac{\omega_0-\omega}{\sqrt{(\omega_0-\omega)^2+\omega_1^2}}.
\end{equation}
For details see, for example, Ekert {\it et al.}~\cite{ekert}. Note that
any deformation of the path of the spin which preserves this solid
angle leaves the phase unchanged. Thus the phase is not affected
by the speed with which the path is traversed; nor is it very
sensitive to random fluctuations about the path. For an
experimental realization of this scheme see, for example, Suter {\it et
al.}~\cite{pines}.

\section{Holonomic conditional phase gate}

Consider to begin with a system of two non-interacting spin-half
particles $S_a$ and $S_b$. In a reference frame aligned with the
static field, the Hamiltonian reads
\begin{equation}
H_0 = \hbar \omega_a S_{az}\otimes \Id_b + \hbar \omega_b
\Id_a\otimes S_{bz},
\end{equation}
where the frequencies $\omega_a/2\pi$ and $\omega_b/2 \pi$ are the
transition frequencies of the two spins and we have used the
scaled Pauli operators $S_i = \sigma_i/2$. (From now on we assume
that $\omega_a$ and $\omega_b$ are very different with $\omega_a >
\omega_b$.)

If the two particles are sufficiently close to each other, they
will interact, creating additional splittings between the energy
levels.  In the case of  two spin-half particles, the magnetic
field of one spin may directly or indirectly affect the energy
levels of the other spin; the energy of the system is increased by
$\pi \hbar J/2$ if the spins are parallel and decreased by $\pi
\hbar J/2$ if the spins are antiparallel. The Hamiltonian of the
system taking into account this interaction reads
\begin{equation}
H=H_0+2\pi\hbar J S_{az}\otimes S_{bz}.
\end{equation}

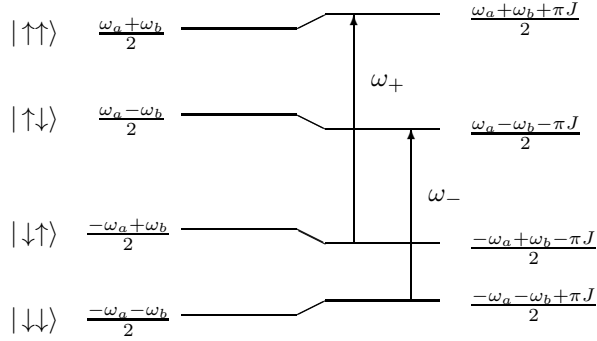
\begin{figure}
\begin{center}
\setlength{\unitlength}{0.03in}
\begin{picture}(100,80)
\put(10,18){$\ket{\downarrow \downarrow}$}
\put(23,18){$\frac{-\omega_a-\omega_b}{2}$}
\put(40,20){\line(1,0){20}} \put(60,20){\line(2,1){5}}
\put(65,22.5){\line(1,0){20}}
\put(90,20.5){$\frac{-\omega_a-\omega_b+\pi J}{2}$}

\put(10,33){$\ket{\downarrow \uparrow}$}
\put(23,33){$\frac{-\omega_a+\omega_b}{2}$}
\put(40,35){\line(1,0){20}} \put(60,35){\line(2,-1){5}}
\put(65,32.5){\line(1,0){20}}
\put(90,30.5){$\frac{-\omega_a+\omega_b-\pi J}{2}$}

\put(10,53){$\ket{\uparrow \downarrow}$}
\put(25,53){$\frac{\omega_a-\omega_b}{2}$}
\put(40,55){\line(1,0){20}} \put(60,55){\line(2,-1){5}}
\put(65,52.5){\line(1,0){20}}
\put(90,50.5){$\frac{\omega_a-\omega_b-\pi J}{2}$}

\put(10,68){$\ket{\uparrow \uparrow}$}
\put(25,68){$\frac{\omega_a+\omega_b}{2}$}
\put(40,70){\line(1,0){20}} \put(60,70){\line(2,1){5}}
\put(65,72.5){\line(1,0){20}}
\put(90,70.5){$\frac{\omega_a+\omega_b+\pi J}{2}$}

\put(70,32.5){\vector(0,1){40}} \put(73,60){$\omega_+$}

\put(80,22.5){\vector(0,1){30}} \put(83,40){$\omega_-$}
\end{picture}
\end{center}
\caption{The energy diagram of two interacting spin-half nuclei.
The transition frequency of the first spin depends on the state of
the second spin.}\label{energydiagram}
\end{figure}

Figure~\ref{energydiagram} illustrates the energy levels of the
system. When spin $S_b$ is in state $\ket{\uparrow}$, the
transition frequency of the spin $S_a$ is
\begin{equation}
\omega_+ = \omega_a + \pi J,
\end{equation}
whereas when spin $S_b$ is in state $\ket{\downarrow}$, the
transition frequency of the spin $S_a$ is
\begin{equation}
\omega_- = \omega_a - \pi J.
\end{equation}

Now suppose that in addition to the static field, we apply a
rotating field that is slowly varied as mentioned in the previous
section. We have seen that the Berry phase acquired by a spin
depends on its transition resonance frequency as given by
Eq.~(\ref{costheta}). Therefore, at the end of a cyclic evolution,
the Berry phase acquired by the spin $S_a$ will be different for
the two possible states of spin $S_b$. Indeed, when spin $S_b$ is
in state $\ket{\uparrow}$, the Berry phase acquired by the spin
$S_a$ is $\gamma_+=\mp\pi(1 - \cos \theta_+)$, with the sign
negative or positive depending on whether spin $S_a$ is up or
down, respectively, and
\begin{equation}
\cos \theta_+ =
\frac{\omega_+-\omega}{\sqrt{(\omega_+-\omega)^2+\omega_1^2}}.
\end{equation}
Similarly, when spin $S_b$ is in state  $\ket{\downarrow}$, the
Berry phase acquired by the spin $S_a$ is $\gamma_-=\mp\pi(1-\cos
\theta_-)$ where
\begin{equation}
\cos \theta_- =
\frac{\omega_--\omega}{\sqrt{(\omega_--\omega)^2+\omega_1^2}}.
\end{equation}

The geometric phase difference $\gamma_+ -\gamma_-$ depends on the 
amplitude of the oscillating magnetic field $\hbar\omega_{1}$ in 
such a way that it has a maximum for a nonvanishing value of $\omega_{1}$. 
Thus, if $\omega_1$ is chosen to be close to this value, fluctuation
errors are of second order and the implementation of the 
conditional phase gate is intrinsically fault tolerant. 

This mechanism effectively implements the conditional phase gate
such as the one demonstrated experimentally by Jones et al using
the NMR technique~\cite{jo}.

\section{Holonomic hadamard gate}
The Hadamard gate requires non-Abelian holonomies. They have been
analyzed from a theoretical point of view in~\cite{zan} using the 
path ordered integration. However,
probably the simplest, experimentally viable construction has
been presented by Bergmann {\it et al.}~\cite{una} and 
Duan {\it et al.}~\cite{duan}. In this construction two
degenerate qubit states $|0\rangle$ and $|1\rangle$ and two
ancilla states $|a\rangle$ and $|b\rangle$ are used together with
an interaction Hamiltonian
\begin{equation}
H=\hbar[|b\rangle(\omega_{0}\langle 0|+\omega_{1}\langle 1|
+\omega_{a}\langle a|)+h.c.]
\end{equation}
with the degenerate eigenvalues $\lambda_{1,2}=0$ and the 
nondegenerate eigenvalues $\lambda_{3,4}=\pm \hbar\Omega$, where
$\Omega=(\omega_{0}^{2}+\omega_{1}^{2}+\omega_{a}^{2})^{1/2}$. If
we parameterize $\omega_0=B \sin\theta\cos\phi$, $\omega_1=B
\sin\theta\sin\phi$, and $\omega_a=B\cos\theta$ the eigenvectors 
for the degenerate eigenvalue read
\begin{eqnarray}
\chi_{1}&=&\sin\phi|0\rangle-\cos\phi|1\rangle \nonumber \\
\chi_{2}&=&\cos\theta\cos\phi|0\rangle+\cos\theta\sin\phi|1\rangle-
\sin\theta|a\rangle.
\end{eqnarray}
A cyclic adiabatic evolution in the parameter space, starting and 
ending at $\phi=0$ and with fixed $\theta$, generates a non-Abelian 
holonomy of the form
\begin{equation}
U=\exp\left[ i\sigma_{y}\int_0^t {\dot \phi}\cos\theta
d\tau\right]=
\left(\begin{array}{cc}
\cos\gamma&-\sin\gamma\\
\sin\gamma&\cos\gamma
\end{array}\right),
\end{equation}
where $\gamma=\int_0^t {\dot \phi}\cos\theta
d\tau$ is given by the swept solid angle in the space of
parameters $\{\theta, \phi\}$. For $\gamma=\pi/4$ the
non-Abelian phase matrix will be the Hadamard gate.

\section{Discussion}
In all experimental realizations, in addition to the geometric
phases there will also be dynamical phases, which depend
on experimental details. In principle these could be calculated
and corrected for using, for example, a conventional spin echo
technique~\cite{jo}. We have implicitly assumed adiabatic schemes
for implementations of the holonomic gates. This does not have to
be the case -- for a nonadiabatic scheme of the conditional see,
for example, Wang {\it et al.}~\cite{wang}.The phase shifts gates, both
the single qubit and the conditional phase shift gates have
been implemented. The Hadamard gate, which requires non-Abelian
holonomies is more difficult to implement by purely geometric
means. However, the Duan {\it et al.} proposal~\cite{duan}
gives hope that it will be implemented in a not too distant future

\section*{Acknowledgments}
I wish to thank Artur Ekert, Bj\"orn Hessmo, and Erik Sj\"oqvist for
discussions. Financial support from the European Science Foundation Quantum
Information Theory programme and the Swedish Research Council
(NFR) is acknowledged.

\end{document}